\documentclass[twocolumn,prl]{revtex4}  

\usepackage{graphicx}
\usepackage{amssymb}

\begin{document}

\def\k {{\bf k}}
\def\n {{\hat{\bf n}}}
\def\r {{\bf r}}
\def\u {{\bf u}}

\def\D {{\bf D}}
\def\G {{\bf G}}
\def\H {{\bf H}}
\def\A {{\bf A}}

\title{Disorder-induced zero-energy spectral  singularity for   
random matrices with correlations}

\author{S.~N.~Taraskin and S.~R.~Elliott} 
\affiliation{Department of Chemistry, University of Cambridge,
             Lensfield Road, Cambridge CB2 1EW, UK}

\date{\today}

\begin{abstract}
A zero-energy mid-band singularity has been found 
in the energy spectrum of random matrices 
with correlations between diagonal and 
off-diagonal elements typical of vibrational 
problems. 
Two representative classes of matrices,  
characterizing  the instantaneous 
configurations in liquids  
and mechanically unstable lattices (which 
mimic the former) have been analysed. 
At least
for  disordered lattice models, the singularity is universal 
and its origin can be explained within the mean-field 
treatment. 

\end{abstract}
\pacs{63.50, 63.20.Dj, 61.20.Gy, 71.23.-k}

\maketitle


Many physical phenomena can be described 
by the Anderson Hamiltonian with off-diagonal 
disorder (see e.g. \cite{Brouwer_00} and references 
therein). 
Stochastic transport \cite{Webman_81},  
atomic vibrations in disordered structures (see e.g. 
\cite{Taraskin_01:PRL} and references therein) and 
instantaneous normal mode (INM) analysis in liquids 
and glasses 
\cite{Keyes_97:review} (called below ''vibrational'' 
problems) are  among these. 
However, there is one key point which 
distinguishes   vibrational problems from the 
standard electron problems: there exist 
strong sum-rule correlations  between 
the off-diagonal and diagonal elements 
(the sum of all the elements for a particular 
row is zero)  of the relevant random matrices  
 for the vibrational problem. 

In different dimensions ($D=1 - 3$), 
the energy spectrum of the electron Anderson 
Hamiltonian with pure off-diagonal disorder 
defined on the simple (hyper) cubic lattice 
(but not, as we have checked, on the f.c.c. lattice) 
exhibits a peculiar feature: a disorder-induced,  
mid-band (zero-energy) singularity occurs 
 \cite{Cain_99,Brouwer_00}. 
Normally, disorder smears out sharp features 
in a spectrum (e.g. van Hove singularities), but 
in this case it creates a  singularity,  
the origin of which in different 
dimensions  is still controversial. 
However, can disorder induce  a similar (or different) 
 singularity   for vibrational problems and, if so, 
can its origin be revealed? 
In this Letter,  
we demonstrate the existence of a mid-band zero-energy singularity 
in the spectrum of  dynamical matrices 
involved in an INM analysis of  
 a model of a monatomic liquid, and investigate 
 analytically and numerically  
the nature of this zero-energy singularity 
in the spectrum of 
random matrices with sum-rule correlations, appearing in disordered 
lattices which mimic  topologically disordered liquids.

\begin{figure} 
\centerline{\includegraphics[width=5.5cm,angle=270]{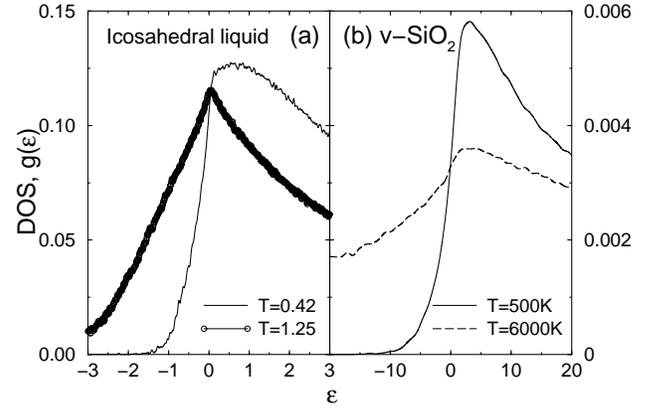}}
\caption{ 
The energy spectrum, $g(\varepsilon)$,   for the instantaneous 
dynamical matrices  for an  INM 
analysis of (a) a liquid with predominantly icosahedral order 
($1620$-particle model averaged 
over $90$ configurations) at two temperatures, as marked 
(the units are the same as in 
Ref.~\protect\cite{Simdyankin_01}) 
and of (b) glassy (the solid curve) 
and liquid (the dashed curve) silica ($1650$-particle model averaged 
over $20$ configurations). The energy units for silica are THz$^2$. 
} 
\label{Sing_INM}
\end{figure} 

We have observed the occurrence of a singularity 
by numerical experiment, in which   structural 
models of a liquid  with predominantly 
icosahedral  order (similar to those discussed 
in Ref.~\cite{Simdyankin_01}) and of 
glassy and liquid silica (similar to those discussed 
in Ref.~\cite{Taraskin_00:PRB_IR1}) have been constructed 
at different temperatures, $T$, by slow molecular 
dynamics quenches (by constant volume and temperature steps). 
The INM spectra of the dynamical matrix 
are shown in Fig.~\ref{Sing_INM}. 
A zero-energy singularity is clearly present in  the energy 
spectrum, $g(\varepsilon)$, for the icosahedral liquid and 
possibly in silica (at least, the tendency of the change 
of the DOS around zero is consistent with the model presented below). 
Similar zero-energy peaks have been found {\it but not explained}  
in liquid theories \cite{Wu_92,Wan_94} and in  simulations 
of a binary Lennard-Jones liquid \cite{Sastry_01}. 
Normally, the density of states (DOS) in the frequency domain, 
$g_\omega= 2\sqrt{|\varepsilon|} g(\varepsilon)$, 
is of interest in the INM analysis, 
but the singularity is masked there 
by the linear factor $2\omega$, 
and that is, probably,  why it has  not been 
 carefully studied before. 

Therefore, there exists a peculiarity in the spectrum 
of  random matrices with sum-rule correlations 
(instantaneous dynamical matrices) and its  
origin can be revealed  analytically as follows.   
Consider a  Hamiltonian describing both the electron 
and vibrational problems: 
\begin{equation}
{\hat \H}= 
\sum_i(\varepsilon_i - \gamma\sum_{j\ne i}t_{ij}
) 
|i\rangle\langle i| + 
\sum_{i,j\ne i}t_{ij}|i\rangle\langle j| 
\ , 
\label{e1}
\end{equation}
where $\varepsilon_i$ and $t_{ij}$ stand for the random 
on-site energies  and the random transfer integrals 
between sites $i$ and $j$, respectively. 
The parameter $\gamma$  
controls the correlations between diagonal 
and off-diagonal matrix elements. 
The standard electron Anderson Hamiltonian with pure off-diagonal 
disorder corresponds to  
$\gamma=0$ and $\varepsilon_i = 0$, with  $t_{ij}$ being   
random variables taken, for example, from a 
 uniform (box)  distribution of width $2\Delta$ centred around 
$t_0=-1$, $t_{ij}\in[t_0-\Delta, t_0+\Delta] $.  
The vibrational (scalar) 
problem corresponds to $\gamma=1$ with the 
 other parameters being the same.  
For atomic vibrations, 
the Hamiltonian corresponds to the dynamical operator, 
transfer integrals to the force constants  
and energy to the squared frequency, 
$\varepsilon =\omega^2$ \cite{Taraskin_01:PRL}.

\begin{figure} 
\centerline{\includegraphics[width=6.cm,angle=270]{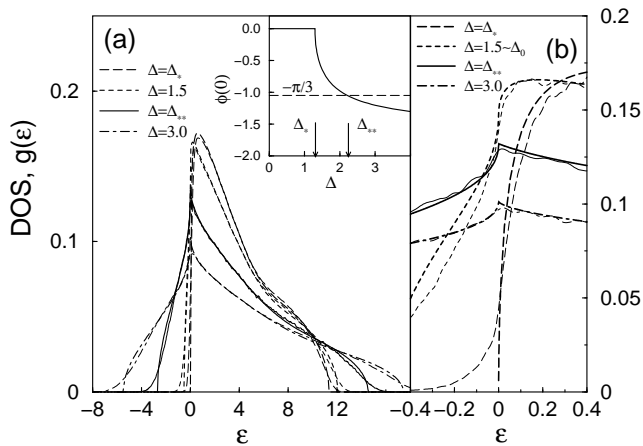}}
\caption{
(a) Evolution of the zero-energy singularity 
in the DOS for a disordered f.c.c. lattice 
with various 
values of disorder, $\Delta$  
($\Delta_*\simeq 1.3$ and $\Delta_{**}\simeq 2.2$). 
The thick curves were obtained by the numerical 
integration of the CPA DOS given by 
Eq.~(\protect\ref{e2}),   
while the thin ones 
 are the results of precise numerical KPM 
solution for vector vibrations in 
a $130\times130\times130$ site f.c.c. lattice. 
The difference between the CPA and KPM curves around 
 both band edges is due to the known failure 
of CPA to reproduce the localized band tails.  
The inset  shows the  
phase, $\phi(0)$, of the effective 
field at zero energy versus disorder, $\Delta$. 
The critical values of disorder are indicated
 by arrows. 
 (b) An enlargement of the zero-energy 
region. 
The CPA results obtained by use of 
Eqs.~(\protect\ref{e2}) and (\protect\ref{e4}) 
are not distinguishable 
on this energy scale. 
} 
\label{Singularity_3D_new}
\end{figure} 

We have calculated by 
the 
kernel polynomial method (KPM) \cite{Silver_97} 
the spectrum of the 
''vibrational'' Hamiltonian (\ref{e1}) 
(see the thin lines in Fig.~\ref{Singularity_3D_new}) 
 for $\gamma=1$ and $\varepsilon_i=0$,  defined 
on the f.c.c. lattice
(actually, of its more general version for vector 
vibrations; see \cite{Taraskin_01:PRL} 
for more detail). 
For sufficiently large degrees of disorder,  
$\Delta > \Delta_0$ ($\Delta_0 \simeq 1.4 - 1.5$), 
the zero-energy singularity is evident 
(Fig.~\ref{Singularity_3D_new}). 
Moreover, the shape of the singularity is similar to that 
found for the topologically disordered models 
(Fig.~\ref{Sing_INM}), indicating its possible universality 
(see below).    
We have also calculated the DOS for the same problem 
within a mean-field approach 
(the single-bond  coherent potential approximation 
(CPA) \cite{Webman_81}), and found 
 remarkably good agreement with the precise numerical (KPM) 
results (cf. thick and thin lines in Fig.~\ref{Singularity_3D_new}) 
in the singularity region for $\Delta \agt \Delta_0$, 
($(\Delta_0-\Delta_*)/\Delta_* \ll 1$, see below 
for a discussion of $\Delta_*$), 
i.e. when the localization threshold  is far  enough  
below zero energy. 
This agreement is surprising (at first sight), 
because it is commonly believed (see e.g. 
\cite{Jarrell_01}) that  mean-field theories 
fail to reproduce  sharp features 
in a spectrum. 
For example, the most successful homomorphic 
cluster CPA \cite{Yonezawa_79,Li_88} well reproduces 
the whole spectrum for the electron problem with 
pure off-diagonal disorder ($\gamma=0$), except for 
the zero-energy singularity. 
Below, we explain why the CPA reproduces the 
zero-energy singularity 
for vibrational problems but not for the electron one,  
and we use this insight to reveal 
the physical origin of the singularity. 

The zero-energy singularity occurs for 
$\gamma=1$ and disappears if $\gamma \ne 1$, 
i.e. when the exact sum-rule correlations of elements in the 
dynamical matrix break down.  
The functional form  of the zero-energy singularity 
observed  is universal 
(i.e. independent of the type of distribution of the force constants,  
 the reference lattice symmetry, 
 scalar or vector 
type of vibrations,  etc.) for the class of  Hamiltonians 
given by Eq.~(\ref{e1}) with  $\gamma=1$ 
and $\varepsilon_i=0$,   
and depends only on the dimensionality of the problem. 
For sufficiently large disorder, the (parity-independent) 
singularity occurs in the mid-band region of the spectrum,  
far from the localized states in the $3D$-case  
(as checked by multifractal analysis)   
and in the range of prelocalized states 
for lower dimensions. 
The analytic mean-field solution shows that  
its appearance is dictated by the universal  non-analytic behaviour 
of the spectral-density operator in the plane-wave 
basis and,  essentially,  is a consequence of the fact that  
zero-energy (long-wavelength) plane waves 
contribute anomalously (but not solely)  to the 
disordered eigenstates with energies 
around zero. 

\begin{figure} 
\centerline{\includegraphics[width=7.cm,angle=270]{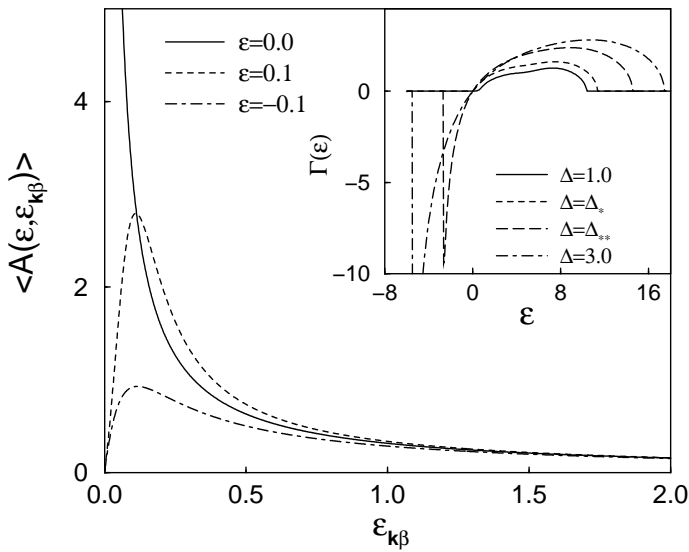}}
\caption{
Dependence of the spectral density, 
 $\langle A(\varepsilon,\varepsilon_{\k\beta}) 
\rangle $, 
on the  crystalline energy, $\varepsilon_{\k\beta}$, 
at values of disordered energy, $\varepsilon$, 
as marked (for $\Delta= \Delta_{**}$),   for 
 vector vibrations in the 
force-constant disordered f.c.c. lattice.  
The inset shows the energy dependence 
of the spectral-density  
peak width parameter, $\Gamma(\varepsilon)$  
(Eq.~(\protect\ref{e3})),  for different 
values of disorder, as marked. 
} 
\label{A_vs_E_cryst}
\end{figure}

The DOS, $g(\varepsilon)$, for a disordered lattice  is the trace 
of the spectral-density operator,  
${\hat \A}(\varepsilon)=\langle \delta(\varepsilon -{\hat\H})\rangle $,   
taken in the convenient orthonormal basis 
of  crystalline eigenstates $|\k,\beta\rangle$ 
($\k$ is the wavevector 
of a plane wave from the branch $\beta$) 
of the same Hamiltonian, but without disorder 
($\Delta =0$), 
and averaged ($\langle\dots\rangle$) 
over the distribution of random variables $t_{ij}$, 
$
 g(\varepsilon) =  
\int \langle 
      \langle 
          \k,\beta|\hat\A(\varepsilon)|\k,\beta
      \rangle
\rangle 
g^{\rm cryst}(\varepsilon_{\k\beta})
{\rm d}\varepsilon_{\k\beta} 
$, 
with 
$ g^{\rm cryst}(\varepsilon_{\k\beta})$ being 
the  crystalline VDOS. 
The diagonal matrix element of the spectral-density 
operator, 
$  A(\varepsilon,\varepsilon_{\k\beta}) \equiv  
\langle \k,\beta|\hat\A(\varepsilon)|\k,\beta\rangle 
$, 
 can be found within the CPA  
\cite{Elliott_74,Ehrenreich_76,Economou_83:book,Taraskin_01:PRL} 
via the complex effective interaction field 
$z(\varepsilon)=z'(\varepsilon)+{\rm i}z''(\varepsilon)$,  
so that 
\begin{equation}
 g(\varepsilon) = 
-\frac{1}{\pi}
\int \frac{
(\varepsilon_{\k\beta}/\varepsilon)
\Gamma(\varepsilon)
} 
{
(\varepsilon_{\k\beta}-{\overline\varepsilon}(\varepsilon))^2 + 
\Gamma^2(\varepsilon) 
}
g^{\rm cryst}(\varepsilon_{\k\beta}){\rm d}\varepsilon_{\k\beta}
\ , 
\label{e2} 
\end{equation}
with 
\begin{equation}
{\overline\varepsilon}(\varepsilon) = 
\varepsilon z'(\varepsilon)/|z(\varepsilon)|^2 
\  \mbox{and} \  
\Gamma(\varepsilon)=
-\varepsilon z''(\varepsilon)/|z(\varepsilon)|^2 
\ .   
\label{e3} 
\end{equation}
The effective interaction field, $z(\varepsilon)$, 
is found from the solution of the self-consistent 
CPA equation (see \cite{Taraskin_01:PRL}). 
The real part, $z'(\varepsilon)$, fluctuates around 
its crystalline value $z_0'=1$, and the imaginary part 
is smooth and negative in the region of 
the non-vanishing disordered 
spectrum, and zero otherwise. 
The  spectral density, 
$ \langle A(\varepsilon,\varepsilon_{\k\beta})\rangle $, 
as a function of $\varepsilon_{\k\beta}$, 
has the shape of a peak, characterized by a width parameter, 
$\Gamma(\varepsilon)$, located at an energy, 
$\varepsilon_{\rm peak}= 
|\varepsilon| |z(\varepsilon)|^{-1} = 
\left[{\overline \varepsilon}(\varepsilon) + 
\Gamma(\varepsilon) \right]^{1/2}$ (see Fig.~\ref{A_vs_E_cryst}). 

As follows from Eq.~(\ref{e2}),  the disordered DOS 
is generically related to the crystalline DOS for the 
reference system convolved with 
the peak-shaped spectral density, 
$\langle A(\varepsilon,\varepsilon_{\k\beta})\rangle$.  
For small disorder ($\Delta \to 0$), the 
spectral density is very narrow and  
close in functional form  to $\delta(\varepsilon - \varepsilon_{\k\beta})$, 
so that the disordered DOS strongly resembles its 
crystalline counterpart. 
With increasing disorder 
but for $\Delta \le \Delta_*$ 
(where the spectrum of the Hamiltonian is still non-negative 
and the lattice is mechanically stable \cite{Taraskin_01:PRL}), 
the spectral-density peak  becomes 
broader and washes out  all the van Hove singularities 
in the crystalline spectrum, except the boundary 
singularity around the lowest band edge, 
$\varepsilon_{\rm min} =0$. 

When the disorder exceeds the critical CPA value, 
$\Delta > \Delta_*$, the number of negative 
force constants becomes so large that the system 
is no longer mechanically stable even in the 
mean-field limit,   and  
negative eigenvalues appear in the spectrum 
(as in instantaneous configurations of liquids  
\cite{Keyes_97:review}), 
i.e. the  lower boundary  of the CPA spectrum 
moves below zero energy, 
$\varepsilon_{\rm min}<0$. 
Exactly in this regime, 
the mid-band zero-energy singularity in the disordered 
CPA spectrum evolves. 
Indeed, if we look at the evolution of the 
spectral-density width parameter, $\Gamma(\varepsilon)$, 
with increasing disorder (see the inset in 
Fig.~\ref{A_vs_E_cryst}), 
we can clearly see 
that the peak width is still zero at 
zero energy, $\Gamma(0)=0$,  
being a consequence of Eq.~(\ref{e3}),  
even though the  value of the effective field 
becomes finite at this point ($z''(0)\ne 0$). 
This means that the spectral density, 
$\langle A(\varepsilon=0,\varepsilon_{\k\beta})\rangle$, as 
a function of the crystalline energy, $\varepsilon_{\k\beta}$, 
has a singularity at $\varepsilon_{\k\beta}=0$ 
(see Fig.~\ref{A_vs_E_cryst}). 
The zero-energy point, in this regime 
($\Delta > \Delta_*$), belongs to the mid-band region.  
The finite value of $z''(0)$ 
 immediately  gives rise to 
 a different singular shape  of the 
spectral density, 
$\langle A(0,\varepsilon_{\k\beta}) 
\rangle   \simeq -\pi^{-1}(z''(0)/|z(0)|^2)\varepsilon_{\k\beta}^{-1}$ 
(see Fig.~\ref{A_vs_E_cryst}),  as 
compared to the 
$\delta$-functional shape of the spectral density 
for the $\Delta < \Delta_*$ regime,  
where both  $\Gamma(0)$ and $z''(0)$ are zero. 
The $\delta$-functional shape of the spectral 
density reproduces the crystalline van Hove singularity, 
$g(\varepsilon)\propto \varepsilon^{(D/2)-1}$,  
at the lower band edge, 
but the $\varepsilon_{\k\beta}^{-1}$-shape of the 
spectral density at $\varepsilon=0$, convolved with 
the crystalline DOS (see Eq.~(\ref{e2})), results 
in a singularity of a different type in the disordered DOS which is not 
related to the crystalline van Hove singularity at 
$\varepsilon = 0$ (see below). 

The shape of the mid-band zero-energy singularity can 
be obtained analytically within the CPA by 
splitting the integration region in Eq.~(\ref{e2}) into a  
low-energy part, 
 $\varepsilon_{\k\beta} < \varepsilon_0$,   where 
$g^{\rm cryst}(\varepsilon) \simeq \chi_{D}
\varepsilon^{(D/2)-1}$ (the Debye law), and the 
rest of the band, $\varepsilon_{\k\beta} > \varepsilon_0$ 
(irrelevant for the shape of the singularity). 
The final result for  
the disordered DOS ($3D$-case) in the limit $|\varepsilon|\to 0$ 
is the following:  
\begin{equation}
g(\varepsilon) \simeq 
-\frac{C\sin\phi(\varepsilon) }{\pi|z(\varepsilon)|} +
\frac{\chi_3|\varepsilon|^{1/2}}{|z(\varepsilon)|^{3/2}} 
\sin \left[
     \frac{3\phi(\varepsilon)}{2} + \theta(\varepsilon)\frac{\pi}{2}
     \right] 
 \ , 
\label{e4} 
\end{equation}
where 
$C=\int\left[g^{\rm cryst}(\varepsilon)/
\varepsilon\right]{\rm d}\varepsilon$ is a model-dependent 
constant,  $\phi(\varepsilon) \equiv 
\arg\left[z(\varepsilon)\right]$ is the phase of 
the effective field 
and $\theta(\varepsilon)$ is the Heavyside function, 
$\theta(\varepsilon<0)=0$ and 
$  \theta(\varepsilon > 0)=1$. 
Bearing in mind that the  effective field  
is a smooth function 
of energy around zero, we can conclude  
from Eq.~(\ref{e4}) that, in the $3D$-case,  
the disordered DOS is a continuous 
function around zero but its derivative shows $\varepsilon^{-1/2}$ 
singular behaviour (see Figs.~\ref{Sing_INM},~\ref{Singularity_3D_new}). 

The expression ~(\ref{e4}) for $g(\varepsilon)$ 
has been obtained in the  limit 
$|\varepsilon| \to 0$ for any degree of disorder. 
The evolution of the singularity with 
disorder (see Fig.~\ref{Singularity_3D_new}) is determined 
by the dependence  on $\Delta$  
of the effective field at zero energy. 
The phase, $\phi(0;\Delta)$, varies in a critical 
manner with disorder 
(see the inset in Fig.~\ref{Singularity_3D_new}(a)),  
being zero below the critical disorder, $\Delta \le \Delta_*$, 
when the zero-energy singularity is just a lower 
band-edge van Hove singularity. 
Above the critical disorder, $\Delta>\Delta_*$, 
the derivative of the DOS  becomes singular 
from both sides about $\varepsilon=0$. 
If the disorder is not too large, 
$\Delta_*<\Delta < \Delta_{**}$, 
the sign of the derivative is positive  on  
 both sides of the singularity and the DOS 
is monotonic in the singularity range  
(see the upper curve in 
Fig.~\ref{Singularity_3D_new}(b)). 
For higher disorder, $\Delta > \Delta_{**}$, 
the derivative changes  sign at the singularity 
and the disordered DOS exhibits a sharp maximum at 
zero energy (see the lower curve in 
Fig.~\ref{Singularity_3D_new}(b)). 
The characteristic value of disorder, 
$\Delta =\Delta_{**}$, at 
which such a transformation of the shape of the singularity 
occurs (the solid line 
in Fig.~\ref{Singularity_3D_new}), 
can be found from the 
solution of the  equation 
 $\phi(0;\Delta_{**})=-\pi/3$ (see the inset in 
Fig.~\ref{Singularity_3D_new}(a)), 
resulting from the condition, $g'(\varepsilon\to 0+)=0$.   

The physical significance of this transition 
is related to the fact that, at $\Delta\simeq\Delta_{**}$, 
the peak width of the spectral density around the 
singularity  becomes comparable 
with the peak position. 
This means that $\Delta_{**}$ corresponds to  the Ioffe-Regel 
crossover for the propagation of plane waves characterized 
by the  energies $\varepsilon_{\k\beta}\to 0$ 
\cite{Taraskin_00:PRB_IR1}. 
In other words, for $\Delta \alt \Delta_{**}$, there is 
a finite low-energy interval of the weak-scattering regime for 
plane-wave propagation, but 
for $\Delta \agt \Delta_{**}$, all plane waves propagate 
in the regime of  strong scattering. 

Similar CPA analyses can be performed for 
lower dimensions, and they result in zero-energy 
singularities as well. 
In $2D$
 (for $\Delta > \Delta_*$), a logarithmic 
singularity, 
$g(|\varepsilon|\to 0) \simeq \pi^{-1}(z''(\varepsilon)/|z(\varepsilon)|^2) 
\chi_2\ln|\varepsilon|$, 
 evolves on the background 
of the van Hove band-edge singularity. 
In $1D$, the divergence of the  disordered DOS
from both sides of zero energy  is even more 
pronounced: 
$g(\varepsilon)\simeq -\chi_1 |\varepsilon|^{-1/2} 
|z(\varepsilon)|^{-1/2}\sin(\phi(\varepsilon)/2-
\theta(\varepsilon)\pi/2)$. 

The link  between  lattice 
models and topologically disordered liquids 
and glasses is, of course, not straightforward. 
The above mean-field analysis for disordered lattices 
 uses the existence of an orthonormal plane-wave basis, 
which is   absent in topologically disordered systems.  
Nevertheless,  an orthogonalized basis  resembling  
a plane-wave basis, at least in the  region of zero energy, 
can readily be 
 constructed \cite{Taraskin_00:PRB_IR1}, 
so that it might be expected  that the same universal 
singular behaviour of the DOS around $\varepsilon=0$ 
should also occur 
for topologically disordered systems.  
Indeed, the singularity is evident for a model liquid 
with predominantly icosahedral order 
(Fig.~\ref{Sing_INM}(a)), 
and it is remarkable that the evolution of the shape of 
this singularity with decreasing temperature mirrors that found 
in the lattice models with decreasing force-constant disorder 
(Fig.~\ref{Singularity_3D_new}(b)). 
Both regimes of the singular behaviour are evident 
from the numerical experiment (Fig.~\ref{Sing_INM}(a)): 
(i) the high-temperature liquid state (circles)
can be characterized by relatively large disorder 
($\Delta \agt \Delta_{**}$);   
(ii) the state just above the glass-transition temperature  
(solid curve) can be associated with the 
intermediate disorder regime ($\Delta_* 
\alt \Delta \alt \Delta_{**}$). 
The situation for vitreous silica 
(Fig.~\ref{Sing_INM}(b)) is different. 
At both temperatures, below and above glass transition, 
 the models stay in the regime 
$\Delta \alt \Delta_{**}$,  which is not surprising 
because silica is a very strong glass-forming liquid  
(especially in comparison with the fragile system 
shown in Fig.~\ref{Sing_INM}(a)), and is characterized 
by well-defined and stable local tetrahedral order even at very 
high temperatures (this means that the disorder, 
i.e. $\Delta$, is relatively small). 
A more detailed analysis (including the glass-transition region) 
 will be presented elsewhere. 

In conclusion, 
we have  demonstrated the presence of 
a mid-band zero-energy singularity in the spectrum 
of the instantaneous dynamical matrices of 
topologically  disordered structural models 
of liquids and of the dynamical matrices of 
disordered lattices which well mimic the former. 
The presence of exact sum-rule correlations 
between the diagonal and off-diagonal elements 
in the disordered dynamical matrix causes this singularity.   
The shape of the disorder-induced 
 singularity, at least in lattice models, 
is universal and depends only 
on the dimensionality of the model. 
Such a universality  is related to the  
universal $\varepsilon^{-1}_{\k\beta}$-behaviour 
of the spectral density in the plane-wave basis,  
which is due to the multiplicative nature of the 
effective interaction mean field (the  mean-field
energy of the quasi-particles, 
${\tilde \varepsilon}_{\k\beta}$,  
is the product of the dimensionless effective 
interaction, $z(\varepsilon)$, 
and the bare crystalline energy, $\varepsilon_{\k\beta}$, i.e 
${\tilde \varepsilon}_{\k\beta} = 
 z(\varepsilon)\varepsilon_{\k\beta}$). 
This property distinguishes the above 
 class of Hamiltonians from those  
 with pure on-site energy disorder, 
characterized by an additive effective field 
\cite{Elliott_74},  which 
do not exhibit a zero-energy singularity. 
Thus, the failure of the homomorphic effective 
field (containing both additive and multiplicative 
contributions) in  reproducing the zero-energy singularity 
for the Anderson Hamiltonian with 
pure off-diagonal disorder without correlations 
in simple cubic lattices can be explained. 
A successful  effective field  for such a problem 
 should  not  contain the additive part, at least at zero energy, 
but its construction is still an open question.



\end{document}